# Superconducting fluctuations probed by the Higgs mode in $Bi_2Sr_2CaCu_2O_{8+x}$ thin films

**Kota Katsumi[1], Zhi Zhong Li[2], Hélène Raffy[2], Yann Gallais[3] and Ryo Shimano[1,4]**

*[1] Department of Physics, The University of Tokyo, Hongo, Tokyo 113-0033, Japan*

*[2] Laboratoire de Physique des Solides (CNRS UMR 8502), Bâtiment 510, Université Paris-Saclay, 91405 Orsay Cedex, France*

*[3] Matériaux et Phénomènes Quantiques UMR CNRS 7162, Bâtiment Condorcet, Université de Paris, 75205 Paris Cedex 13, France*

*[4] Cryogenic Research Center, The University of Tokyo, Yayoi, Tokyo 113-0032, Japan*

## Abstract

Superconducting (SC) fluctuations in cuprate superconductors have been extensively studied to gain a deep insight into preformed Cooper pairs above the SC transition temperature $T_c$. While the various measurements, such as the terahertz (THz) optical conductivity, Nernst effect, angle-resolved photoemission spectroscopy (ARPES), and scanning tunneling microscopy (STM) measurements have provided the signature of the SC fluctuations, the onset temperature of the SC fluctuations depends on the measurement scheme. Here, we shed light on the Higgs mode to investigate the SC fluctuations, as it is the direct fingerprint of SC order parameter and can help elucidate the development of SC phase coherence. We perform THz pump-optical probe spectroscopy for underdoped and overdoped $Bi_2Sr_2CaCu_2O_{8+x}$ (Bi2212) thin films. The oscillatory behavior in the pump-probe signal (THz Kerr signal) observed in the SC phase has been identified as the Higgs mode in single crystals in our previous work [K. Katsumi *et al.*, Phys. Rev. Lett. **120**, 117001 (2018)], but two onset temperatures are identified above $T_c$. Combined with the results of the single crystals in a wide range of doping, we find that the first onset $T_1^{ons}$ is 10-30 K above $T_c$. $T_1^{ons}$ coincides with that of the superfluid density $N_s$ extracted from the THz optical conductivity. Hence, $T_1^{ons}$ is interpreted as the onset of macroscopic SC phase stiffness. On the other hand, the second onset $T_2^{ons}$ is identified at substantially higher than $T_c$, whose origin is discussed in terms of preformed Cooper pairs.

# I.    INTRODUCTION

Among the various physical properties in cuprate superconductors, superconducting (SC) fluctuations have attracted intensive interests over decades [1] and been extensively investigated to uncover the Cooper pairing above $T_c$ by various experimental approaches: terahertz (THz) [2–5] and microwave spectroscopy [6–8], infrared spectroscopy [9–11], torque magnetometry [12–15], Nernst [16,17], angle-resolved photo emission spectroscopy (ARPES) [18–21], scanning tunneling microscopy (STM) [22] measurements and ultrafast pump-probe spectroscopy [23–26]. Most of these experiments have investigated the onset temperature of the SC fluctuations ($T_{Onset}$) to understand how the SC coherence emerges from the complex metallic state, yet $T_{Onset}$ depends on experimental techniques and a unified picture on the onset of SC fluctuations is still lacking. For instance, in the THz and microwave spectroscopy the onset temperature $T_{Onset}$ has been identified at 10-20 K above $T_c$ in $Bi_2Sr_2CaCu_2O_{8+x}$ (Bi2212) [2,3], $La_{2-x}Sr_xCuO_4$ (LSCO) [4–6] and $YBa_2Cu_3O_{7-x}$ (YBCO) [8]. On the contrary, the Nernst and ARPES and infrared spectroscopy measurements assert much higher temperatures of $T_{Onset}$, up to 50-100 K above $T_c$ [9–11,16–21]. However, the Nernst signal was shown to be enhanced by the stripe order and its origin remains under debate [27]. Therefore, alternative measurements which are sensitive to the SC order parameter are called for.

To gain deeper insights into the onset of SC fluctuations, we focus on the collective amplitude mode of the SC order parameter, namely the Higgs mode, which directly manifests the development of the SC order parameter [28–30]. For decades, it has been considered that the Higgs mode in superconductors is invisible except for some compounds that exhibit the charge-density wave (CDW) and superconductivity [31–35]. However, with the advances of nonlinear THz spectroscopy [36,37], it has become possible to observe the Higgs mode by pump-probe measurements and THz third-harmonic generation (THG), as demonstrated first in an $s$-wave superconductor NbN [38–40]. In parallel, substantial theoretical progress has been made on the understanding of observability of Higgs mode in the nonlinear THz optical responses, and the importance of paramagnetic coupling of the light-matter interaction has been elucidated [30,41–49].

More recently, the Higgs mode has been observed in the $d$-wave cuprate superconductor Bi2212 by THz pump-optical probe (TPOP) spectroscopy in reflection geometry. Phenomenologically, the measurement scheme resorts to the THz pulse-induced Kerr effect, where the intense THz pulse modifies the optical response of the near-infrared probe pulse. The oscillatory behavior of the optical reflectivity, which follows the squared THz-pump electric field ($E$ field) was identified as a fingerprint of Higgs mode [50]. Subsequently, THG from cuprate superconductors has also been observed in other cuprate superconductors such as LSCO, YBCO and DyBa$_2$Cu$_3$O$_{7-x}$ and interpreted in terms of Higgs mode-mediated THG [51]. Nevertheless, the onset temperature of the Higgs mode in cuprates has not been clarified.

In this paper, we investigate the SC fluctuations in underdoped (UD) and overdoped (OD) Bi2212 thin films by TPOP measurements. The instantaneous oscillatory component, or THz Kerr signal, observed in the pump-probe signal has been identified as the Higgs-mode response in our previous study with single-crystalline Bi2212 samples [50]. We find two onset temperatures in the THz Kerr signal; the first onset $T_1^{ons}$ which is about 10 K above $T_c$, and the second onset $T_2^{ons}$ which is located far above $T_c$. $T_1^{ons}$ coincides with that of the superfluid density $N_s$, evaluated by the THz complex optical conductivity measurement on the same samples. This indicates that $T_1^{ons}$ can be attributed to the onset of dynamically fluctuating SC phase stiffness, below which the Higgs mode becomes visible within the picosecond snapshot of the THz probe pulse. The results are also compared with that of Bi2212 single-crystalline samples with various doping, revealing that the dynamically fluctuating SC phase stiffness in Bi2212 extends up to 10-30 K above $T_c$ over a wide range of carrier doping. Furthermore, $T_2^{ons}$ shows a good agreement with the SC gap opening temperature reported from the STM measurements.

## II. EXPERIMENT

We performed TPOP measurements on underdoped (UD76, $T_c$ = 76 K and thickness of 60 nm) and overdoped (OD67, $T_c$ = 67 K and thickness of 160 nm) Bi2212 thin films grown by the sputtering method on MgO substrates. The SC transition temperature $T_c$ was determined by the magnetic susceptibility measurement using a superconducting quantum interference device (SQUID) as shown in Appendix A. In Fig. 1(a) we show a schematic of the TPOP spectroscopy. The output of a regenerative amplified Ti:sapphire laser system with the pulse duration of 100 fs, central wavelength of 800 nm, pulse energy of 4 mJ and repetition rate of 1 kHz was divided into two beams: one for the generation of the intense THz-pump pulse and the other for the optical probe pulse. To generate an intense monocycle THz pulse, we employed the tilted-pulse-front technique with a LiNbO$_3$ crystal [52] combined with the tight focusing method [53]. The THz-pump $E$ field was detected by electro-optic sampling in a 380-$\mu$m GaP (110) crystal placed inside the cryostat. The waveform and power spectrum of the THz-pump pulse are represented in Figs. 1(c) and 1(d), respectively. The central frequency of the THz-pump $E$ field is located around 0.6 THz = 2.4 meV, which is much smaller than the anti-nodal SC gap energy, $2\Delta_0 > 20$ meV for the present doping levels of the Bi2212 samples [54–57]. We used the peak $E$ field of 130 kV/cm for UD76 and 220 kV/cm for OD67 in order to ensure that $\Delta R/R$ is within the third-order nonlinear regime where it scales as the square of the THz-pump $E$ field. The TPOP measurements were performed as a function of both the pump and probe polarization angles $\theta_{\text{Pump}}$, $\theta_{\text{Probe}}$ as defined in Fig. 1(b) to discriminate the Higgs-mode and other contributions (for details of the methodology, see Ref. [50]). Since the oscillatory component of the reflectivity change follows the square of the THz peak $E$ field [Fig. 2(e)], the amplitude of the oscillatory component i.e. the THz Kerr signal can be expressed by the third-order nonlinear susceptibility $\chi^{(3)}$ as [50,58]

$$\frac{\Delta R}{R}(E_i^{\text{Probe}}, E_j^{\text{Probe}}) = \frac{1}{R}\frac{\partial R}{\partial \varepsilon_1}\varepsilon_0 \text{Re}\chi_{ijkl}^{(3)}E_k^{\text{Pump}}E_l^{\text{Pump}}, \tag{1}$$

where $\varepsilon_1$ is the real part of the dielectric constant and $E_i$ denotes the $i$-th component of the THz-pump or optical-probe $E$ field. Assuming the tetragonal symmetry for Bi2212, the nonlinear susceptibility can be decomposed into the irreducible representations of the $D_{4h}$ point group as [50,59]

$$\chi^{(3)}(\theta_{\text{Pump}},\theta_{\text{Probe}}) = \frac{1}{2}(\chi_{\text{A}_{1g}}^{(3)} + \chi_{\text{B}_{1g}}^{(3)}\cos2\theta_{\text{Pump}}\cos2\theta_{\text{Probe}} + \chi_{\text{B}_{2g}}^{(3)}\sin2\theta_{\text{Pump}}\sin2\theta_{\text{Probe}}). \tag{2}$$

Here we defined $\chi_{\text{A}_{1g}}^{(3)} = \chi_{xxxx}^{(3)} + \chi_{xyyy}^{(3)}$, $\chi_{\text{B}_{1g}}^{(3)} = \chi_{xxxx}^{(3)} - \chi_{xyyy}^{(3)}$ and $\chi_{\text{B}_{2g}}^{(3)} = \chi_{xyxy}^{(3)} + \chi_{xyyx}^{(3)}$. In our previous work, mean-field calculations showed that the Higgs-mode response should appear only in the isotropic $\text{A}_{1g}$ channel [50]. Figure 2(f) shows the probe polarization dependence of the reflectivity change $\Delta R/R$ for OD67 at 15 K and demonstrates that $\Delta R/R$ is dominated by the isotropic $\text{A}_{1g}$ component although a slight angle dependence is discerned, which is consistent with the results of the Bi2212 single crystals [50]. Therefore, we attribute the observed THz Kerr signal below $T_c$ to the Higgs mode, and we focus on the $\text{A}_{1g}$ components of $\Delta R/R$ in the following discussion.

In addition to the TPOP measurement, THz time-domain spectroscopy (THz-TDS) in transmission geometry was performed on the same UD76 and OD67 thin films to evaluate the superfluid density, which corresponds to the phase stiffness of the SC order parameter.

## III.   RESULT

### A.  THz pump-optical probe spectroscopy

The $\text{A}_{1g}$ component of the THz pulse-induced transient reflectivity change $\Delta R/R$ for $\theta_{\text{Pump}} = 45°$ at various temperatures is displayed in Fig. 2(a) for UD76 and Fig. 2(b) for OD67 thin films, respectively. At 30 K, below $T_c$, $\Delta R/R$ for both films show oscillatory behaviors which follow the squared THz-pump $E$ field $E_{\text{Pump}}(t)^2$. In addition to the THz Kerr component, $\Delta R/R$ has a decaying component that survives for as long as 10 ps. For the UD76 thin films at 120 K, above $T_c$, the signal consists of a weaker THz Kerr component and a decaying signal that switches its sign after 2 ps. At 268 K the decaying signal remains positive for all delays. For the OD67 thin films at 150 K, above $T_c$, the signal consists of a weaker THz Kerr component and a decaying signal. The overall behaviors are similar to the results obtained in the single crystals [50].

Figures 2(c) and 2(d) display the temperature dependence of the $A_{1g}$ components of the transient reflectivity change $\Delta R/R$ for UD76 and OD67, respectively. In both samples $\Delta R/R$ displays a sharp increase from slightly above $T_c$.

As the Higgs-mode oscillation is expected to follow $E_{Pump}(t)^2$, we can extract the amplitude of the THz Kerr signal from the fast Fourier transformation (FFT) of $\Delta R/R$. Figures 3(a) and 3(b) show the FFT spectrum of $E_{Pump}(t)^2$ and the $A_{1g}$ component of $\Delta R/R$ for thin films at selected temperatures. The FFT of the $A_{1g}$ component of $\Delta R/R$ around 1.5 THz, which corresponds to the peak in the FFT of the squared THz-pump $E$ field, increases as the temperature is lowered. The FFT amplitudes for UD76 and OD67 integrated from 1.2 to 2.2 THz ($A_{FFT}$) are shown in Figs. 3(c) and 3(d), respectively. With decreasing temperature, the integrated FFT amplitude $A_{FFT}$ sharply increases below 100 K in both samples, while it shows a more gradual increase at higher temperatures. The origin of $A_{FFT}$ at higher temperature will be discussed later in this section.

To quantitatively discuss the temperature dependence of the amplitude of the THz Kerr signal, we should take into account the temperature dependence of the squared THz-pump $E$ field inside the thin film in the analysis of third-order nonlinear susceptibility $\chi^{(3)}$ relevant to the THz Kerr signal. Here, we evaluated the squared THz-pump $E$ field inside the thin film ($B_{FFT}$) for each temperature by using the optical constants measured by THz transmission spectroscopy. The detailed calculations of $B_{FFT}$ and $\chi^{(3)}$ are explained in Appendix B. The obtained temperature dependence of $B_{FFT}$ are shown in Figs. 4(a) and 4(b). The third-order nonlinear susceptibility $\chi^{(3)}$ is calculated by dividing the integrated FFT amplitude $A_{FFT}$ by $B_{FFT}$. Figures 4(c) and 4(d) show the resulting third-order nonlinear susceptibility $\chi^{(3)}$ as a function of temperature for UD76 and OD67, respectively; it gradually increases as the temperature decreases from 200 K, and shows an upturn at about 10 K above $T_c$. As our previous work has demonstrated [50], we attribute the $A_{1g}$ component of the THz Kerr signal below $T_c$ to the Higgs mode.

In order to define an onset temperature of the THz Kerr signal ($T_1^{ons}$) precisely, we take the second derivative of $\chi^{(3)}$ with respect to temperature as shown in Figs. 4(c) and 4(d). The onset temperature $T_1^{ons}$ is determined as the slope change in the temperature dependence of the second derivative. We extract the onset temperature of the THz Kerr signal, $T_1^{ons} = 90$ K for UD76 and $T_1^{ons} = 80$ K for OD67. It is tempting to attribute the finite $\chi^{(3)}$ response at $T_c < T < T_1^{ons}$ to SC fluctuations, where the long-range phase coherence is not statically established but emerges in a short time scale, as introduced to explain previous THz spectroscopy results [2–5]. This interpretation will be further examined in the next section by comparing $T_1^{\text{ons}}$ with the onset temperature of the phase stiffness extracted from linear THz spectroscopy results on the same samples. We also identify a second onset temperature of $\chi^{(3)}$ ($T_2^{\text{ons}}$) from the slope change in the temperature dependence of $\chi^{(3)}$, $T_2^{\text{ons}} = 185$ K for UD76 and 116 K for OD67, as shown in the insets of Figs. 4(c) and 4(d), which is substantially higher than $T_c$. A similar second onset temperature is identified for the case of single-crystalline samples with various doping. The origin of this second onset temperature $T_2^{\text{ons}}$ will be argued in the discussion section. Note that the $\chi^{(3)}$ signal above $T_2^{\text{ons}}$ sustains even at room temperature for most of the samples. This high temperature signal can be attributed to rather generic nonlinear transport in the normal metal phase as observed in metallic film and particles [60–66].

## B. THz transmission spectroscopy

In order to compare the onset temperature of the THz Kerr signal $T_1^{\text{ons}}$ and that of the SC phase stiffness, we evaluate the superfluid density by THz-TDS in transmission geometry. Figures 5(a)-(d) show the real and imaginary parts of the THz optical conductivity for UD76 and OD67 samples. Below $T_c$, the imaginary part of the optical conductivity $\sigma_2(\omega)$ for both thin films exhibit $1/\omega$-like divergent behavior which is a signature of SC condensation. The real and imaginary parts of the optical conductivity are reasonably fitted simultaneously by the two-fluid model, which is given by [67]

$$\sigma_1(\omega) = \frac{\omega_p^2 \tau}{1+\omega^2\tau^2} + N_s\delta(\omega),$$
$$\sigma_2(\omega) = \frac{\omega_p^2\tau^2\omega}{1+\omega^2\tau^2} + \frac{N_s}{\omega}. \tag{3}$$

Here, the first term is the Drude component and the second term is the SC component, where $\omega_p$ is the plasma frequency, $\tau$ is the scattering time and $N_s$ is the superfluid density. The previous THz optical conductivity measurements and theories showed that the inhomogeneity transfers the spectral weight from the Delta function to the Drude component [68]. While two Drude components were assumed to describe the optical conductivity in Ref. [68], here only one Drude component is assumed to reduce the number of the fitting parameters. The fitting results are shown in Figs. 5(a)-(d) by solid curves and well reproduce the complex optical conductivity of both thin films. The temperature dependence of the three parameters $N_s$, $\omega_p$ and $\tau$ are plotted in Figs. 5(e)-(h). The superfluid density $N_s$ shown in Figs. 5 (e) and 5(g) sharply increases from above $T_c$ for both films. The growth of superfluid density above $T_c$ is also identified as a low-frequency upturn in the $\sigma_2(\omega)$ spectra with the Drude component subtracted (see

Appendix C for details). The onset temperature of $N_s$ is extracted as $T_{Ns}$ = 90 K for UD76 and $T_{Ns}$ = 80 K for OD67. For UD76 the onset temperature of $N_s$ is consistent with that reported by the previous THz optical conductivity measurement [2]. The scattering time $\tau$ of the Drude component plotted in Figs. 5(f) and (h) gradually increases with decreasing temperature, whose behavior is also consistent with the previous report [68]. While we set the plasma frequency $\omega_p$ as a fitting parameter, it does not show significant temperature dependence for either thin film.

## IV. DISCUSSION

In both Bi2212 thin films, $T_1^{\text{ons}}$ coincides with $T_{Ns}$ within our experimental error bars, which further supports the interpretation that $T_1^{\text{ons}}$ is attributed to the onset of the Higgs mode and therefore that of the macroscopic SC phase coherence. Now we examine $T_1^{\text{ons}}$ of the Bi2212 single-crystalline samples in a wide range of doping from our previous TPOP results [1]. Since the temperature dependence of the THz $E$ field inside the sample does not strongly depend on hole concentration as shown in Figs. 4(a) and 4(b), we evaluated the temperature dependence of the third-order nonlinear susceptibility $\chi^{(3)}$ for single crystals by approximating the temperature dependences of the internal THz $E$ field with that of the UD76 thin film (Appendix D). Figure 6 summarizes $T_1^{\text{ons}}$ for two thin films and five differently doped single crystals. The onset temperature of the Higgs mode $T_1^{\text{ons}}$ is located at 10-30 K above the $T_c$ dome. The result is also consistent with the previous THz conductivity measurements in UD Bi2212 thin films [2,3]. Such a behavior has been reported in other cuprate superconductors including LSCO and YBCO, where the onset temperature of the long-range SC phase coherence is shown to exist at most 20 K above $T_c$ in a wide range of doping when probed by the microwave and THz spectroscopy [4–6,8].

Next, we consider the origin of the second onset temperature of the THz Kerr signal, $T_2^{\text{ons}}$. The values of $T_2^{\text{ons}}$ for all the measured samples including the single crystals are represented in the phase diagram of Fig. 6. It is worth noting here that, except for OD66, $T_2^{\text{ons}}$ shows a good agreement with the local gap opening temperature observed in STM measurements which was ascribed to local SC patches [22]. This coincidence may suggest that $T_2^{\text{ons}}$ is related to the preformed Cooper pairs associated with the gap opening in local SC patches. Previous studies from Nernst [16,17] and ARPES [19–21] measurements in Bi2212 also show the SC gap opening temperature substantially higher than $T_c$, although the values are distributed between $T_1^{\text{ons}}$ and $T_2^{\text{ons}}$ depending on the measurement scheme. In other cuprates such as LSCO and YBCO, SC gap opening temperatures as high as 100 K above $T_c$ have been reported by infrared spectroscopy [9–11] and Nernst measurements [16,17]. Even the two temperature scales, $T_1^{\text{ons}}$ and $T_2^{\text{ons}}$, have also been identified in the Fourier-transform infrared (FTIR) study of YBCO [10], which is consistent with our results on Bi2212 although the materials are different. Here we note that the growth of local SC patches below $T_2^{ons}$ should be observed in principle in the THz-TDS measurements like the previous FTIR studies for YBCO [10-12], whereas it is absent in the present results for Bi2212. This difference may be attributed to the sensitivity of the measurement. In the previous FTIR spectroscopy [11], $N_s$ above $T_1^{ons}$ is estimated as 0.2-0.3% of that at the lowest temperature, while the error bars for $N_s$ in our THz optical conductivity is 5-10% of the value at the lowest temperature. Therefore, although the value of $N_s$ above $T_1^{ons}$ cannot be directly compared between Bi2212 and YBCO, a finite $N_s$ above $T_1^{ons}$ may be below the noise floor in our THz-TDS measurement.

Furthermore, recent nonlinear conductivity, paraconductivity and torque magnetometry measurements have provided evidence that the macroscopic SC phase coherence vanishes rapidly above $T_c$ in an exponential fashion. In these studies the SC phase locking mechanism among the locally formed SC patches is explained by a phenomenological percolation model [15,69,70]. These observations suggest that the origin of the THz Kerr signal between $T_1^{ons}$ and $T_2^{ons}$ deserves further investigation and calls for a microscopic theory of the THz Kerr effect in such a temperature region of precursor superconductivity.

## V. CONCLUSION

We have investigated the Higgs-mode response through the observation of the THz Kerr effect in UD and OD Bi2212 thin films using the TPOP spectroscopy. Two onset temperatures are identified in the THz Kerr signal. Combined with the result of single-crystalline samples, the first one ($T_1^{ons}$) is identified at 10 K-30 K above $T_c$, while the second one ($T_2^{ons}$) is located substantially higher than $T_c$. $T_1^{ons}$ coincides with that of the superfluid density evaluated from the THz optical conductivity measurement, indicating that $T_1^{ons}$ is attributed to the onset temperature of the long-range SC phase coherence which is dynamically fluctuating and snapshotted by the THz probe. On the other hand, the coincidence between $T_2^{ons}$ and the gap opening temperature reported from Nernst [17], ARPES [19] and STM [22] measurements may suggest that $T_2^{ons}$ is related to the development of preformed Cooper pairs. The measurement method presented here would give deeper insights into the equilibrium state, providing access for the study of dynamical interplay between the SC and other competing or coexisting orders in unconventional superconductors. Being a direct probe of the SC order parameter with a picosecond time resolution, the study of the Higgs-mode response would also lay the foundation to investigate the nonequilibrium phenomena, e.g. for the light-induced superconductivity [71–75].

## ACKNOWLEDGMENTS

We acknowledge N. Tsuji for fruitful discussion. This work was supported in part by JSPS KAKENHI (Grants No. 18H05324 and No. 15H02102), Mitsubishi Foundation and JST CREST (Grant No. JPMJCR19T3). K.K. was supported by JSPS Research Fellowship for Young Scientists (Grant No. 19J12873).

## Appendix A: Determination of $T_c$

Figures 7(a) and 7(b) show the magnetic moments for UD76 and OD67 Bi2212 thin films measured by SQUID under zero-field cooling (ZFC) and field-cooling (FC). The SC transition temperature $T_c$ is determined by the onset of drop in the magnetic moment. For the Bi2212 single crystals in our previous work [50], $T_c$ is determined by the same manner.

## Appendix B: Calculation of the third-order nonlinear susceptibility

We evaluated the third-order nonlinear susceptibility by calculating the THz-pump $E$ field inside the thin film as follows. The FFT of the $A_{1g}$ component of $\Delta R/R(t)$ can be described as

$$\frac{\Delta R_{A_{1g}}}{R}(\omega) = \frac{1}{R}\frac{\partial R}{\partial \varepsilon_1}\varepsilon_0 \int_0^\infty dt\, \mathrm{Re}\chi_{A_{1g}}^{(3)} E_{\mathrm{Film}}(t)^2 e^{-i\omega t} = \frac{1}{R}\frac{\partial R}{\partial \varepsilon_1}\varepsilon_0 \mathrm{Re}\chi_{A_{1g}}^{(3)} B(\omega), \tag{B1}$$

where $E_{\mathrm{Film}}(t)$ is the $E$ field inside the thin film in the time-domain and $B(\omega)$ is the FFT of $E_{\mathrm{Film}}(t)^2$. As discussed in our previous work, the third-order nonlinear susceptibility of the Higgs mode does not depend on frequency when the probe photon energy is much higher than the SC gap energy and pump photon energy [50]. Therefore, $\mathrm{Re}\chi_{A_{1g}}^{(3)}$ in Eq. (B1) can be singled out from the integral.

The THz-pump $E$ field inside the film in the time domain $E_{\mathrm{Film}}(t)$ can be calculated by using its FFT $E_{\mathrm{Film}}(\omega)$, which is expressed as [76]

$$\frac{E_{\mathrm{Film}}(\omega)}{E_{\mathrm{Pump}}(\omega)} = \frac{\frac{2}{1+n_{\mathrm{Film}}(\omega)}e^{i(n_{\mathrm{Film}}(\omega)-1)\omega d/c}}{1-\frac{n_{\mathrm{Film}}(\omega)-1}{n_{\mathrm{Film}}(\omega)+1}\frac{n_{\mathrm{Film}}(\omega)-n_{\mathrm{Sub}}(\omega)}{n_{\mathrm{Film}}(\omega)+n_{\mathrm{Sub}}(\omega)}e^{2in_{\mathrm{Film}}(\omega)\omega d/c}}, \tag{B2}$$

where $E_{\mathrm{Pump}}(\omega)$ is the incident THz-pump $E$ field, $d$ is the thickness of the thin film, and $n_{\mathrm{Film}}(\omega)$ and $n_{\mathrm{Sub}}(\omega)$ are the complex refractive index of the thin film and the substrate, respectively. As the THz-pump FFT intensity $E_{\mathrm{Pump}}(\omega)^2$ covers from 0.1 to 3 THz [Fig. 1(d)], we can calculate $E_{\mathrm{Film}}(\omega)$ by using $n_{\mathrm{Film}}(\omega)$ and $n_{\mathrm{Sub}}(\omega)$ in the same frequency region. Because the sample size is $3 \times 3$ mm and not large enough to measure the optical conductivity below 0.4 THz, we used $n_{\text{Film}}(\omega)$ obtained from the fitting to the complex optical conductivity $\sigma(\omega)$ with Eq. (3).

By integrating Eq. (B1), the nonlinear susceptibility can be written as

$$\text{Re}\chi_{\text{A}_{1g}}^{(3)} = \frac{\int_{2\pi*1.2}^{2\pi*2.2} d\omega \frac{\Delta R_{\text{A}_{1g}}}{R}(\omega)}{\frac{1}{R}\frac{\partial R}{\partial\varepsilon_1}\varepsilon_0 \int_{2\pi*1.2}^{2\pi*2.2} d\omega B(\omega)} = \frac{A_{\text{FFT}}}{B_{\text{FFT}}}, \tag{B3}$$

where $A_{\text{FFT}}$ is the integrated amplitudes of $\Delta R/R(\omega)$ from $\omega/2\pi = 1.2$ to 2.2 THz and shown in Figs. 3(c) and 3(d). $B_{\text{FFT}}$ is the integrated amplitudes of $B(\omega)$ from $\omega/2\pi = 1.2$ to 2.2 THz and is shown in Figs. 4(a) and 4(b). Finally, the calculated $\text{Re}\chi_{\text{A}_{1g}}^{(3)} \equiv \chi^{(3)}$ is plotted as a function of temperature in Figs. 4(c) and 4(d).

## Appendix C: Alternative estimation of the onset temperature of superfluid density from $\sigma_2(\omega)$

The signature of superfluid density in $\sigma_2(\omega)$ can be revealed by subtracting the Drude component (the first term of $\sigma_2(\omega)$ in Eq. (3)). We first fit $\sigma_1(\omega)$ spectrum with the Drude model, given by the first term of $\sigma_1(\omega)$ in Eq. (3), as shown by the solid curves in Figs. 8(a) and 8(b). Then we subtracted the corresponding Drude component from the experimentally obtained $\sigma_2(\omega)$. The resultant spectra $\Delta\sigma_2(\omega)$ are represented in Figs. 8(c) and 8(d). An upturn behavior toward the lower frequency range is clearly identified even at 90 K for UD76 and at 70 K for OD67, which indicates the presence of a superfluid component at those temperatures. By plotting the temperature dependence of $\Delta\sigma_2(\omega)$ at the lowest frequency of 1.9 meV for UD76 and that of 2.4 meV for OD67 [Figs. 8(e) and 8(f)], we estimated the onset temperature as $T_{Ns} = 90\pm5$ K for UD76 and $T_{Ns} = 75\pm5$ K for OD67. These values agree well with those extracted by the two-fluid model, which further reinforce our determination of $T_{Ns}$. In the subtracted spectra of $\Delta\sigma_2(\omega)$, we also recognize a finite peak around 8 meV, which is not reproduced by the two-fluid model. This high frequency deviation may be attributed to the non-Drude behavior of cuprate superconductors.

## Appendix D: Temperature dependence of the normalized THz Kerr signal for Bi2212 single crystals

Temperature dependence of the third-order nonlinear susceptibility of the THz Kerr signal $\chi^{(3)}$ for Bi2212 single crystals are summarized in Fig. 9. In all of the single crystals, $\chi^{(3)}$ increases from 10 to 30 K above $T_c$. The onset temperature of the Higgs mode is defined as the temperature where the second derivative of $\chi^{(3)}$ with respect to temperature exhibits a significant deviation from the smooth normal state signal. In all of the single crystals, $T_1^{\text{ons}}$ is located 10-30 K above $T_c$ as summarized in the phase diagram (Fig. 6).

**Figures**

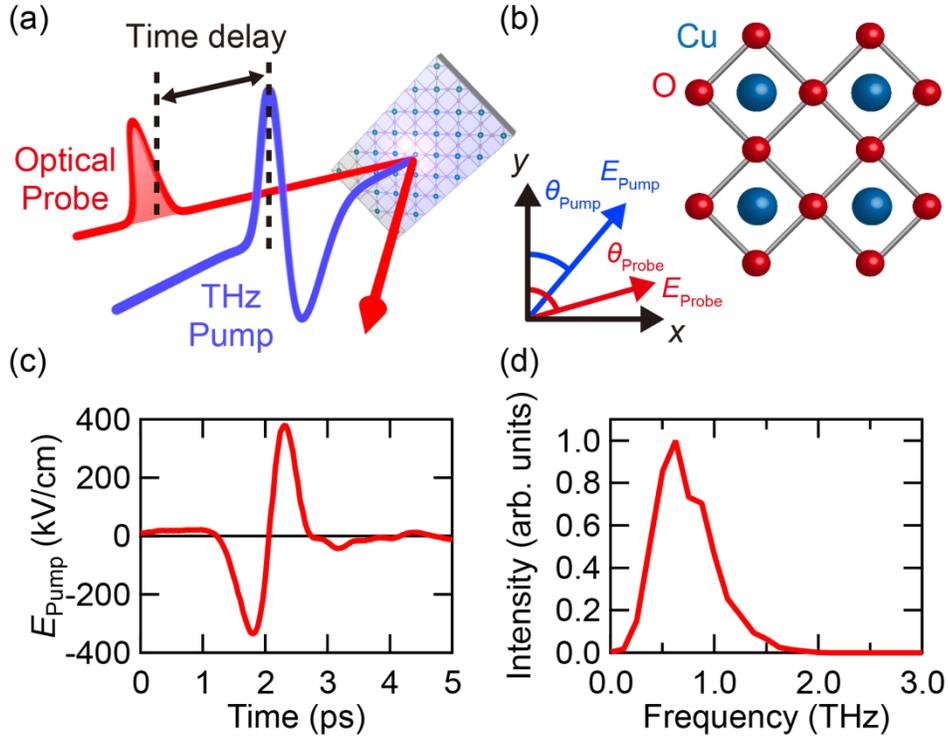

FIG. 1. (a) A geometry for the THz pump-optical probe spectroscopy. The THz-pump pulse and optical-probe pulse propagate collinearly. (b) A schematic illustration of the CuO$_2$ plane. The pump ($\theta_{\text{Pump}}$) and probe ($\theta_{\text{Probe}}$) polarization angles are defined relative to the Cu-O bond ($y$-axis). (c) The waveform of the THz-pump $E$ field. (d) The power spectrum of the THz-pump $E$ field.

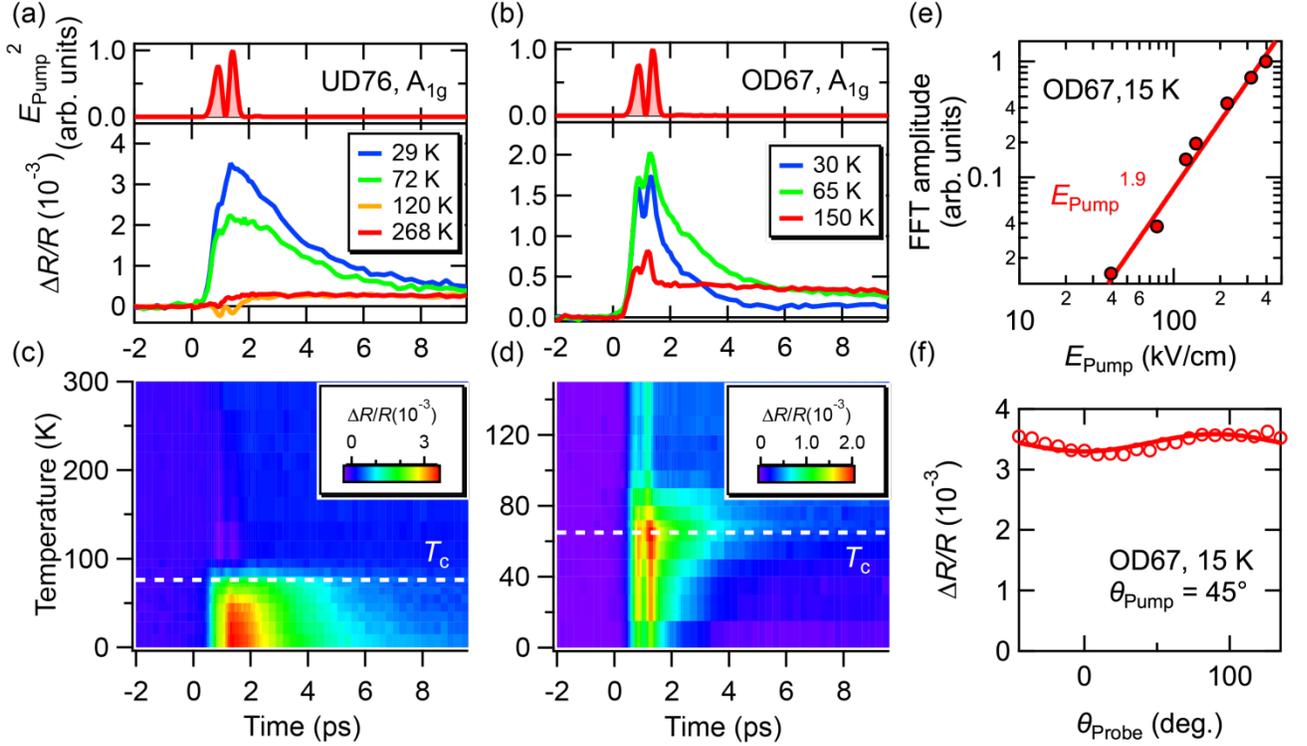

FIG. 2. (a), (b) The A$_{1g}$ components of the THz-pump-induced reflectivity change $\Delta R/R$, at selected temperatures for UD76 and OD67, respectively. The upper panels show the waveforms of the squared THz-pump $E$ field. (c), (d) Temperature dependence of the A$_{1g}$ components of the reflectivity change $\Delta R/R$, for UD76 and OD67, respectively. (e) The FFT amplitude of $\Delta R/R$ at 15 K integrated from 1.2 to 2.2 THz at $\theta_{Pump} = \theta_{Probe} = 45°$ as a function of the peak THz-pump $E$ field for OD67. (f) Probe polarization dependence of $\Delta R/R$ at delay time of 1.3 ps for OD67 when the pump polarization is fixed at $\theta_{Pump} = 45°$.

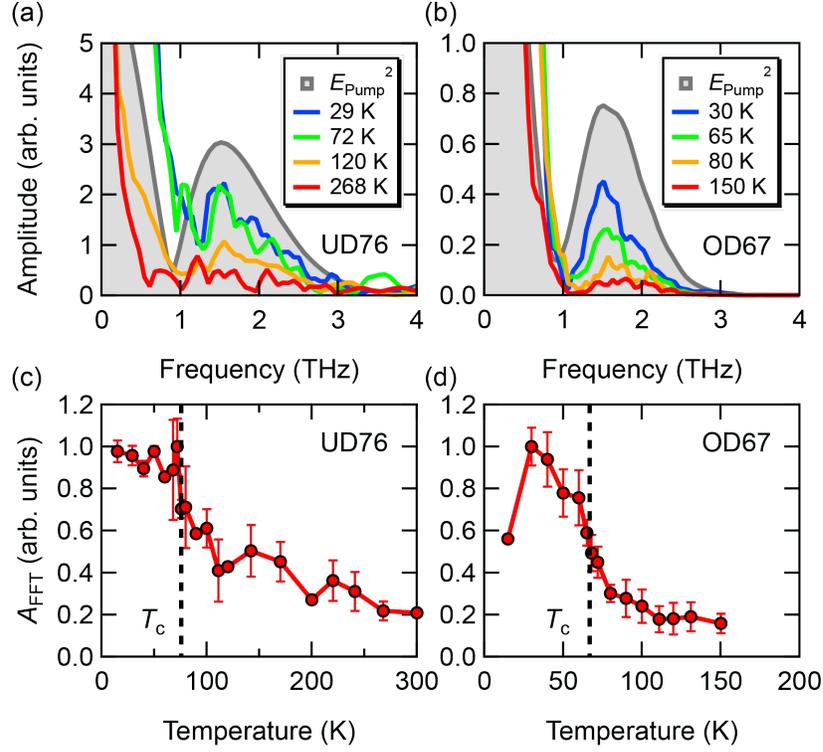

FIG. 3. (a), (b) The FFT amplitude of the $A_{1g}$ component of $\Delta R/R$ at selected temperatures for UD76 and OD67, respectively. The gray curve is the FFT of the squared THz-pump $E$ field ($E_{Pump}^2$). (c), (d) Temperature dependence of the FFT amplitude of $\Delta R/R$ integrated from 1.2 to 2.2 THz ($A_{FFT}$) for UD76 and OD67, respectively.

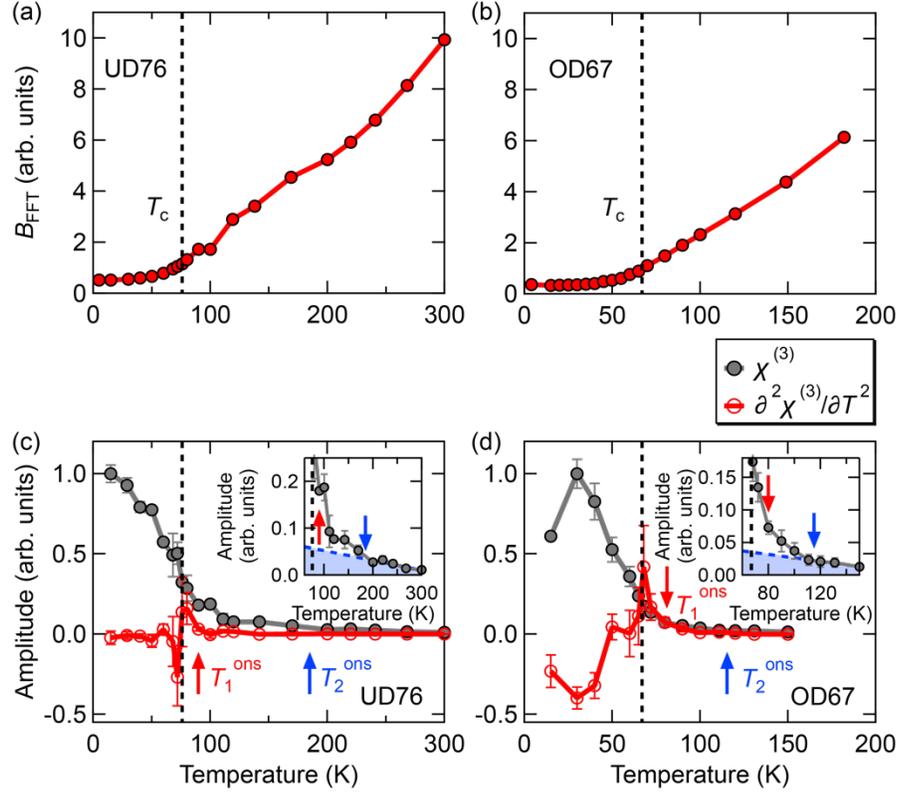

FIG. 4. (a), (b) Temperature dependence of the FFT amplitude of the squared THz-pump $E$ field inside the thin film ($B_{FFT}$) for UD76 and OD67, respectively. The FFT amplitude is normalized by its value at $T_c$. (c), (d) Temperature dependence of the third-order nonlinear susceptibility of the THz Kerr signal $\chi^{(3)}$ in TPOP measurements (the gray curves). The red curves show the second derivative of $\chi^{(3)}$ with respect to temperature. The red vertical arrows denote the onset temperature of the sharp increase in the THz Kerr signals $T_1^{ons}$ determined from the second derivative. The blue vertical arrows denote the onset temperature of the THz Kerr signals $T_2^{ons}$. The insets are the expanded figures around $T_2^{ons}$.

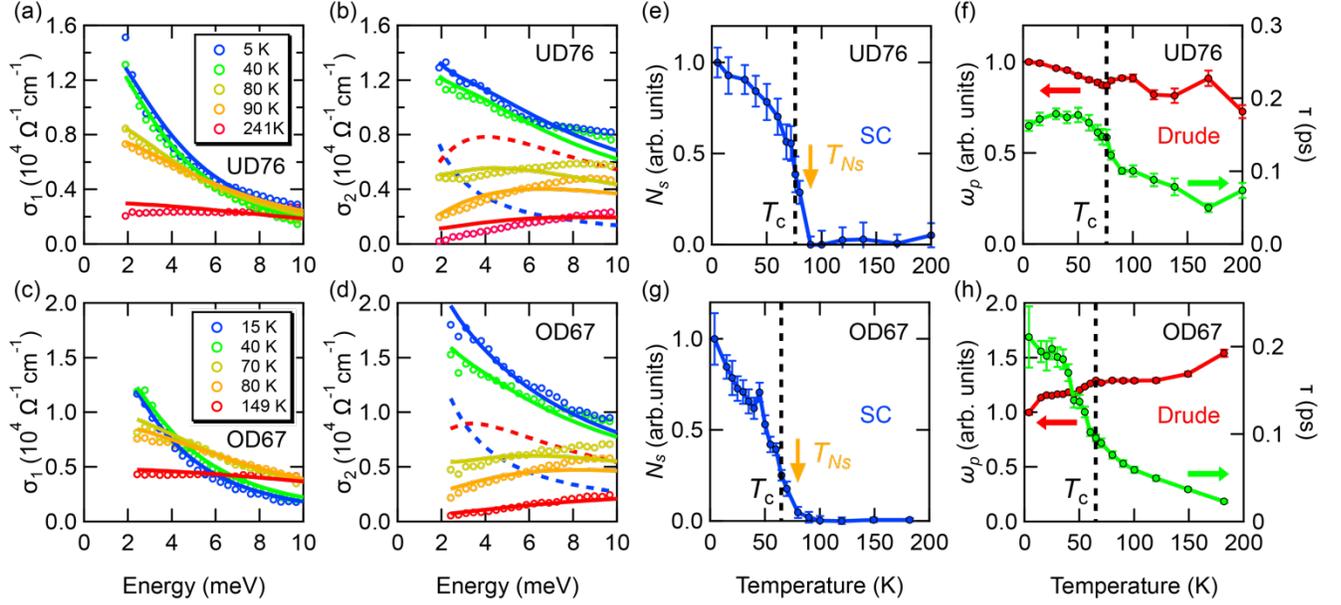

FIG. 5. (a)-(d) Real and imaginary parts of the optical conductivity measured by THz-TDS for UD76 and OD67. The open circles represent the data and the solid lines are the fitting curves by the two-fluid model. In (b) and (d), the SC component (blue) and Drude component (red) at 5 K for UD76 and at 15 K for OD67 are plotted by the dashed curves, respectively. (e)-(h) Temperature dependence of the fitting parameters in Eq. (3) for UD76 and OD67, respectively. (e) and (g) show the temperature dependence of the superfluid density $N_s$. (f) and (h) show the temperature dependence of the plasma frequency $\omega_p$ (left axis) and the scattering time $\tau$ (right axis). Here, $N_s$ and $\omega_p$ are normalized by their value at the lowest temperature. The orange vertical arrows denote the determined onset temperature of $N_s$ ($T_{Ns}$).

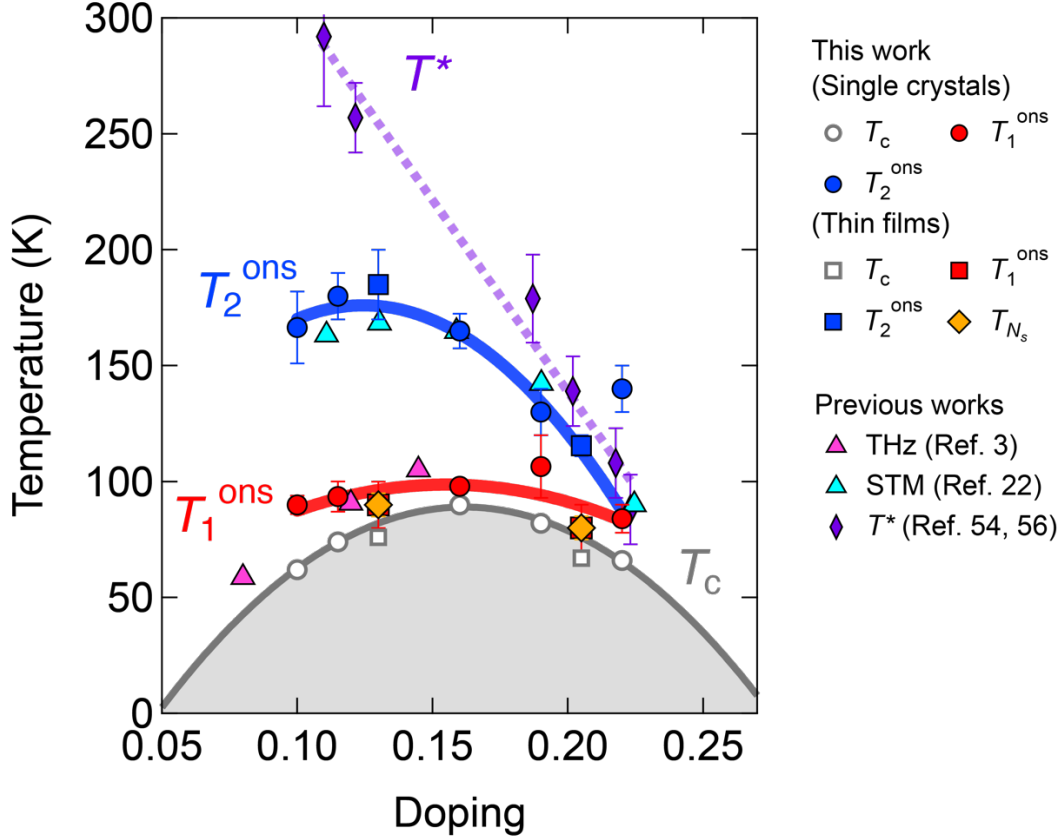

FIG. 6. Doping dependence of the onset temperature of the THz Kerr signal obtained from TPOP experiments and superfluid density evaluated by THz-TDS measurements. The hole concentration is determined from $T_c$ using the Presland and Tallon's equation [77]. The red and blue circles are $T_1^{ons}$ and $T_2^{ons}$ for Bi2212 single crystals that is evaluated from the data in Ref. [50] and the red and blue squares are $T_1^{ons}$ and $T_2^{ons}$ for Bi2212 thin films (this work). $T_{Ns}$ obtained for Bi2212 thin films are shown by orange diamonds (this work). The data of $T_{Ns}$ for other dopings shown by magenta triangles are adopted from Ref. [3]. The light blue triangles are the SC gap opening temperature for Bi2212 adopted from Ref. [22]. The purple diamonds are the pseudogap opening temperature $T^*$ for Bi2212 from Ref. [54,56].

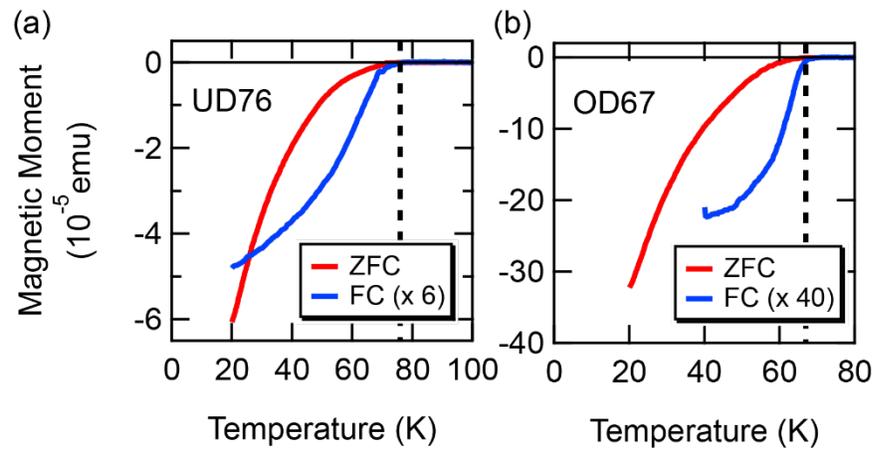

FIG. 7. (a), (b) The magnetic moment of the UD76 and OD67 Bi2212 thin films, respectively. The black dotted lines denote the determined $T_c$.

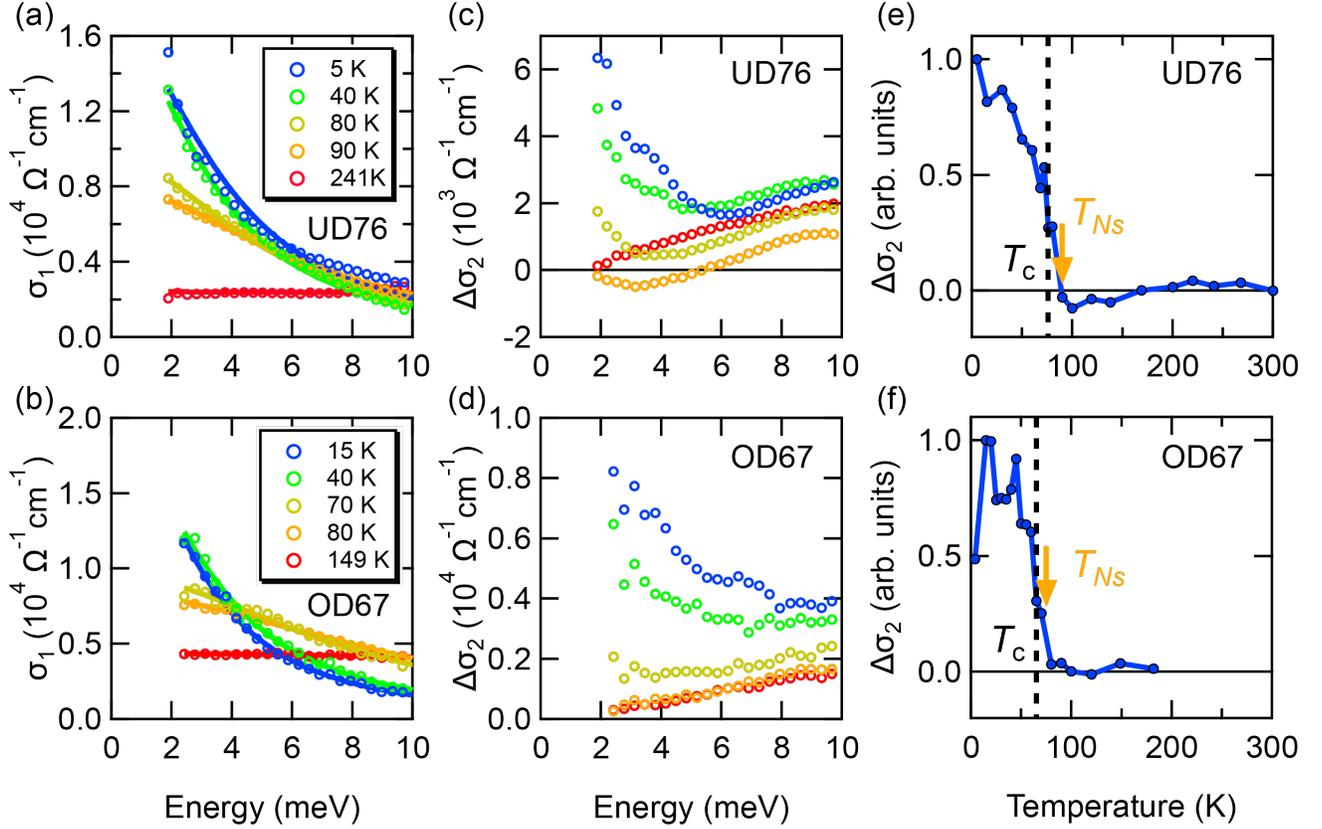

FIG. 8. (a), (b) Real part of the optical conductivity measured by THz-TDS for UD76 and OD67, respectively. The open circles represent the data and the solid lines are the fitting curves by the Drude model. (c), (d) The difference between the imaginary part of the optical conductivity and the Drude component obtained from the fits to the real part. (e), (f) Temperature dependence of $\Delta\sigma_2(\omega)$ at the lowest frequency. The orange vertical arrows denote the determined onset temperature of $\Delta\sigma_2(\omega)$ ($T_{Ns}$).

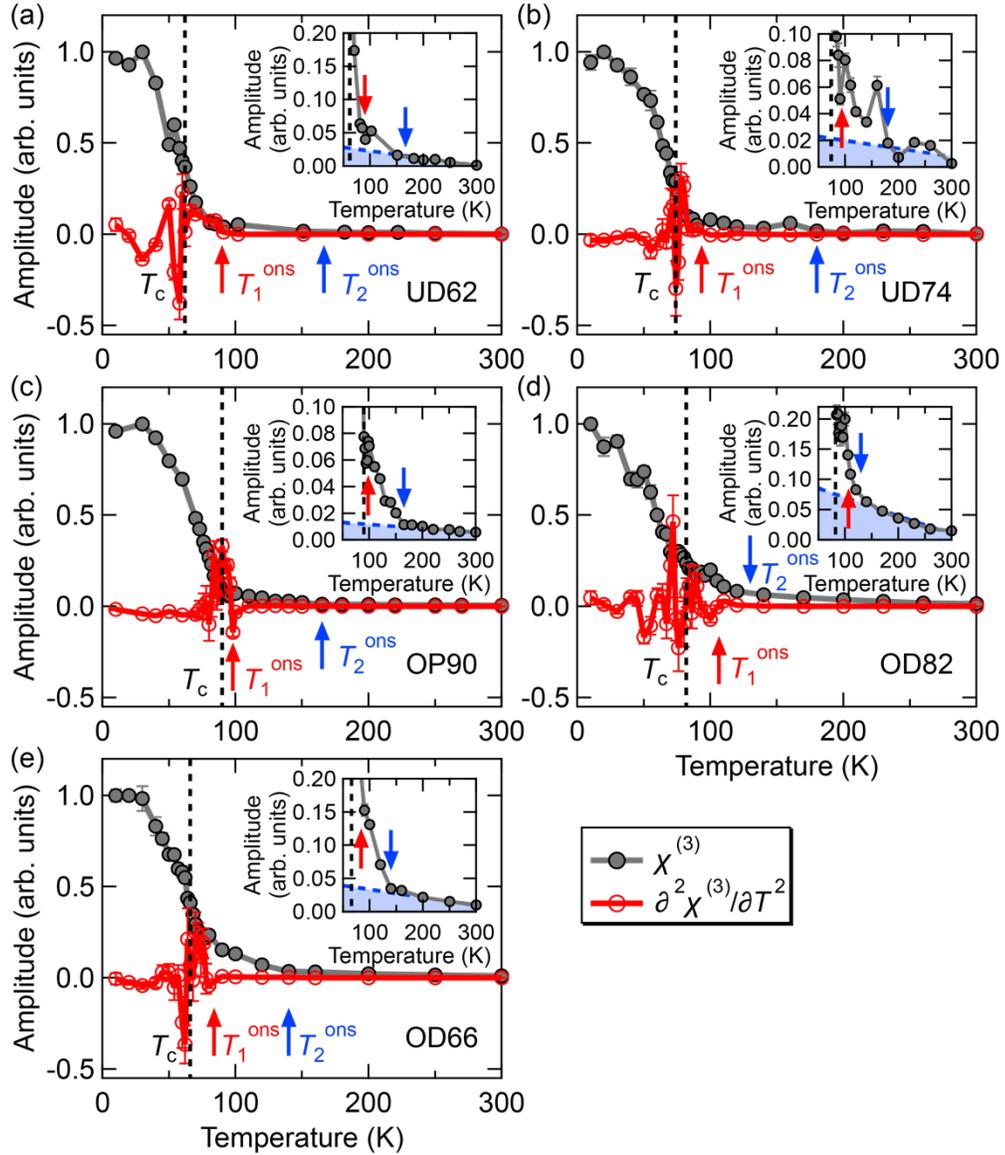

FIG. 9. Temperature dependence of the third-order nonlinear susceptibility of the THz Kerr signal $\chi^{(3)}$ in TPOP measurements for Bi2212 single crystals (the gray curves). The red curves show the second derivative of $\chi^{(3)}$ with respect to temperature. The red vertical arrows denote the onset temperature of the sharp increase in the THz Kerr signals $T_1^{ons}$ determined from the second derivative. The blue vertical arrows denote the onset temperature of the THz Kerr signals $T_2^{ons}$. The insets are the expanded figures around $T_2^{ons}$.